# Transonic Galactic Outflows and Their Influences to the Chemical Evolution of Galaxies and Intergalactic Space

Asuka Igarashi[a], Masao Mori[a] and Shin-ya Nitta[b]

[a]University of Tsukuba, 1-1-1, Tennodai, Tsukuba, Ibaraki, 305-8577, Japan
[b]Tsukuba University of Technology, 4-3-15, Amakubo, Tsukuba, Ibaraki, 305-8520, Japan

**Abstract.** Galactic winds are widely recognized as important ingredients in galaxy evolution, and they impact the chemical enrichment of galaxies and the intergalactic medium. We investigate the acceleration process of isothermal, spherically symmetric steady galactic outflows in an appropriate galactic gravitational potential applying the transonic analysis which is the well-known approach for the solar wind. The results show that the transonic solutions of the galactic winds critically depend upon the mass distribution in a galaxy such as the dark matter halo (DMH) and the central super massive black hole (SMBH). We discover the existence of two types of transonic solutions in the gravity from the combination of DMH and SMBH. The first one is accelerated near the SMBH which is similar to the Parker solution, and the other is slowly accelerated over the entire region of DMH. These two transonic solutions have different mass fluxes and starting points. Therefore, they have different influences to the chemical evolution of galaxies and intergalactic space. We have found that the mass fluxes of two transonic solutions are considerably different by several orders of magnitude in spite of the same mass distribution. This result indicates that mass flux is very sensitive not only to the mass distribution but also to the chosen transonic solution.

## INTRODUCTION

Recent observational cosmology reveals that the primary ingredient of the galactic mass consists the cold dark matter, and it has an essential role affecting not only to the evolution of galaxies but also to the large-scale structure of the universe. The study of galaxy formation in the gravitational potential of the dark matter halo (DMH) indicates that galactic outflows play significant roles on the evolution of galaxies and the metal enrichment of the intergalactic space[1,2]. In order to realize galactic outflows, it needs sufficient thermal energy supply to escape from the gravitational potential well of the galaxy. So far, the study mainly assumed supernovae and stellar winds as the thermal energy source [3,4].

Recently, Tsuchiya et al. (2013) studied the transonic galactic outflow based on the solar wind model assuming the outflow driven by the thermal energy of the interstellar medium itself almost at the virial temperature[5]. Tsuchiya et al. (2013) extended the Parker's model[6] and explored transonic galactic outflows in a realistic gravitational potential of DMH mass distribution under isothermal, spherically symmetric and steady assumption. They used various models of DMH expected observationally and theoretically. As the result, they found that a transonic point exists in a far distance. In previous studies, it is believed that the galactic outflow becomes supersonic in the vicinity with star-forming activity. Therefore, the slowly accelerated galactic outflow which was proposed by Tsuchiya et al. (2013) is a new picture of galactic outflows.

Recent observational studies reveal that most galaxies include a central super-massive black hole (SMBH). It may affect the acceleration process of the outflow in the galactic central region. In this study, we extend Tsuchiya's model of the transonic galactic outflow including the SMBH contribution to reproduce a realistic gravitational potential in the central region. With the gravitational potentials of DMH and SMBH, we have found two different types of transonic solution for the same parameters. These solutions have different starting points and different mass fluxes. Therefore, they give different contributions to the chemical evolution of galaxies and the intergalactic space. We estimate the mass fluxes and their influences to the chemical evolution of the intergalactic space.

# TRANSONIC ANALYSIS

## Model for Galactic Outflows

We assume isothermal, spherically symmetric and steady outflows without mass injection along flow except the starting point. The basic equations are the conservation of mass and momentum as follows,

$$4\pi \rho v r^2 = \dot{M}, \quad \ldots (1)$$

$$v\frac{dv}{dr} = -\frac{c_s^2}{\rho}\frac{d\rho}{dr} - \frac{d\Phi}{dr}, \quad \ldots (2)$$

where $\rho$, $v$, $r$, $\dot{M}$, $c_s$ and $\Phi$ are gas density, gas velocity, radius from the galactic center, mass flux, sound speed and the gravitational potential, respectively. Note that $\dot{M}$ and $c_s$ are constants. Eliminating $\rho$ from equations (1) and (2), we obtain

$$\left(1 - \frac{1}{M^2}\right)\frac{dM^2}{dx} = N(x), \quad \ldots (3)$$

$$N(x) = \frac{4}{x} - \frac{2}{c_s^2}\frac{d\Phi}{dx}, \quad \ldots (4)$$

where $M = v/c_s$ is the Mach number and $x = r/r_d$ is non-dimensional radius with $r_d$ is the scale radius of DMH. Tsuchiya et al. (2013) adopted the model of the density profile of DMH as

$$\rho_{DMH}(r;\alpha) = \frac{\rho_d r_d^3}{r^\alpha (r + r_d)^{3-\alpha}}, \quad \ldots (5)$$

where $\rho_d$ represents the scale density. With $\alpha = 1$, this density profile corresponds to the NFW model[8]. The density is proportional to $r^{-1}$ in the limit of $r \to 0$ in this model. Integrating the equation (3) and combined with equation (5), we obtain

$$M^2 - \log M^2 = 4\log x - 4\Phi'(\alpha, K_{DMH}, K_{BH}; x) + C, \quad \ldots (6)$$

where

$$\Phi'(\alpha, K_{DMH}, K_{BH}; x) = \frac{1}{c_s^2}\Phi = K_{DMH}\int \frac{1}{x^2}\left(\int_0^x x^{2-\alpha}(x+1)^{\alpha-3}dx\right)dx - \frac{K_{BH}}{x}, \quad \ldots (7)$$

$$K_{DMH} = \frac{2\pi G \rho_d r_d^2}{c_s^2}, \quad \ldots (8)$$

$$K_{BH} = \frac{GM_{BH}}{2r_d c_s^2}, \quad \ldots (9)$$

$G$ is the Newton's constant, $M_{BH}$ is mass of SMBH, $C$ is the integration constant. The parameters $K_{DMH}$ and $K_{BH}$ approximately represent the weight of the influence by DMH mass and SMBH mass, respectively.

We summarize transonic solutions in Figure 1 (for $\alpha=1$). Transonic solutions for other $\alpha$ are also shown in Figure 2. We find that transonic solutions are categorized into two cases: A) single X-point and B) two X-points with single O-point. In case B), the transonic solution through the inner X-point is referred to as type $X_{in}$ and the transonic solution through the outer one is referred to as type $X_{out}$ in the followings. The inner X-point is formed by the gravitational potential of SMBH, while the outer one is formed by that of DMH. In case B-1 (see Figure 1), type $X_{in}$ solution starts from the center, but type $X_{out}$ one does not start from center. In case B-2, type $X_{out}$ solution extends to infinity, but type $X_{in}$ one does not extend to infinity. In B-1 and B-2, two transonic solutions have different mass fluxes and starting points. So, we may expect different influences by $X_{out}$ and $X_{in}$ on the star-formation history and the mass of the gas transported to the intergalactic space.

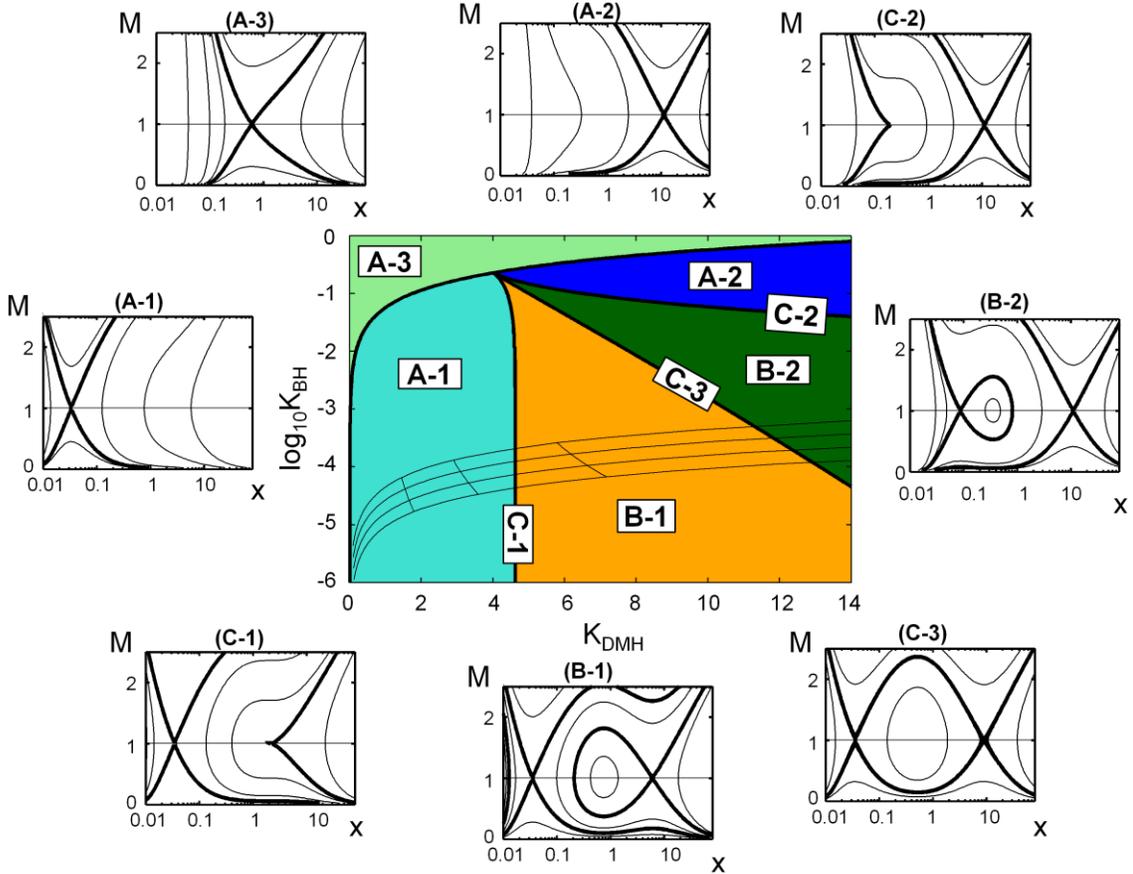

**FIGURE 1.** Various solutions in the gravitational potential of DMH ($\alpha=1$) and SMBH. The horizontal axis is $K_{DMH}$ defined in equation (8) and the vertical axis is $K_{BH}$ defined in equation (9). The labels, such as A-1, represent the types of transonic solutions. See Table 1. for details of these solutions. The four thin solid lines in the central diagram represent the parameter range of actual galaxies from $10^{11} M_\odot$ (bottom) to $10^{14} M_\odot$ (top). The three thin solid lines intersecting four solid lines represent $\eta=0.5$, 1 and 2 from left to right. $\eta$ is defined in the last section (Parameters in Actual Galaxies).

## DISCUSSION

### Parameters in Actual Galaxies

In actual galaxies, values of parameters ($K_{DMH}, K_{BH}$) should be in a plausible range. Using the virial temperature and results of other studies[15,16], we estimate the range of these parameters as

$$K_{DMH} = \eta \frac{c}{2} \left( \int_0^c x^{2-\alpha}(x+1)^{\alpha-3} dx \right)^{-1}, \quad \ldots (13)$$

$$K_{BH} = \frac{0.11\eta c}{2 \times 10^4} \left( \frac{c}{9.7} \right)^{\frac{1.27-1}{-0.074}}, \quad \ldots (14)$$

where $c$ is the concentration parameter defined as the ratio between the virial radius of DMH and $r_d$, and $\eta$ is a fudge factor of the order of unity. We show these actual range of ($K_{DMH}, K_{BH}$) in Figure 1. Each of the four long curves extending horizontally in Figure 1 corresponds to different DMH mass. Each of the three curves intersecting these four lines corresponds to the different temperature. From the plausible parameter range covers a region in $K_{DMH}$-$K_{BH}$ plane, we can result that the expected types of the galactic outflow will be A-1, C-1 and B-1.

| **Table 1.** The features of solutions with the gravitational potential of DMH and central black hole. | | |
|---|---|---|
| A | 1 | Single X-point generated by the central super massive black hole (SMBH). |
|   | 2 | Single X-point generated by the dark matter halo (DMH). |
|   | 3 | Single X-point with no extreme points in N(x). |
| B | 1 | Two X-points with one O-point. The transonic solution, through X-point generated by the DMH, rounds O-point and breaks off. |
|   | 2 | Two X-points with one O-point. The transonic solution, through X-point generated by the SMBH, rounds O-point and breaks off. |
| C | 1 | Boundary solution between A-1 and B-1. |
|   | 2 | Boundary solution between A-2 and B-2. |
|   | 3 | Boundary solution between B-1 and B-2. |

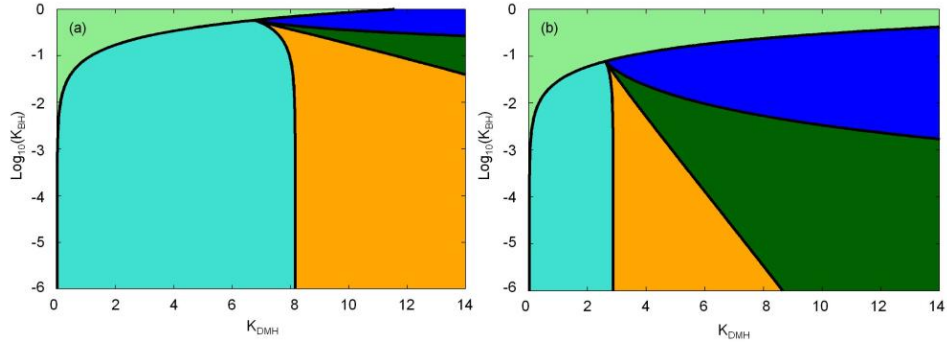

**FIGURE 2.** Various solutions for (a): $\alpha = 0$ and (b): $\alpha = 1.5$. Each color corresponds to the same in Figure 1.

## Chemical Evolution of the Intergalactic Space

The mass flux of the galactic outflow is important to the evolution of galaxies and the release of heavy elements to the intergalactic space. In this section, we estimate the mass fluxes predicted by our model. If there are two transonic solutions (B-1, see Figure 1), the mass fluxes are different between two solutions. The mass fluxes are determined by $K_{DMH}$, $K_{BH}$ and $\rho_0$, where $\rho_0$ is the boundary condition of gas density distribution. Here, we focus on the ratio of mass fluxes of two transonic solutions with the assumption that the gas density distributions of two transonic solutions in subsonic region are nearly similar. We assume the gas density distributions are hydrostatic-like in the subsonic region. The ratio of mass fluxes reduces

$$\xi = \dot{M}_{X_{in}}/\dot{M}_{X_{out}}$$
$$= \rho_{HS}(x_{X_{in}})x_{X_{in}}^2 / \rho_{HS}(x_{X_{out}})x_{X_{out}}^2 \quad \ldots (10)$$

where $x_{X_{in}}$ and $x_{X_{out}}$ are the locus of the inner X-point and that of the outer X-point, respectively. The gas densities $\rho_{HS}(x_{X_{in}})$ and $\rho_{HS}(x_{X_{out}})$ are the hydrostatic density at $x_{X_{in}}$ and that at $x_{X_{out}}$, respectively. The ratio of mass flues $\xi$ is determined by two parameters, $K_{DMH}$ and $K_{BH}$. We show the result in Figure 3 and find that the mass fluxes of two transonic solutions are different by several orders of magnitude in spite of the same mass distribution. This difference should influence to the metal enrichment of the intergalactic space.

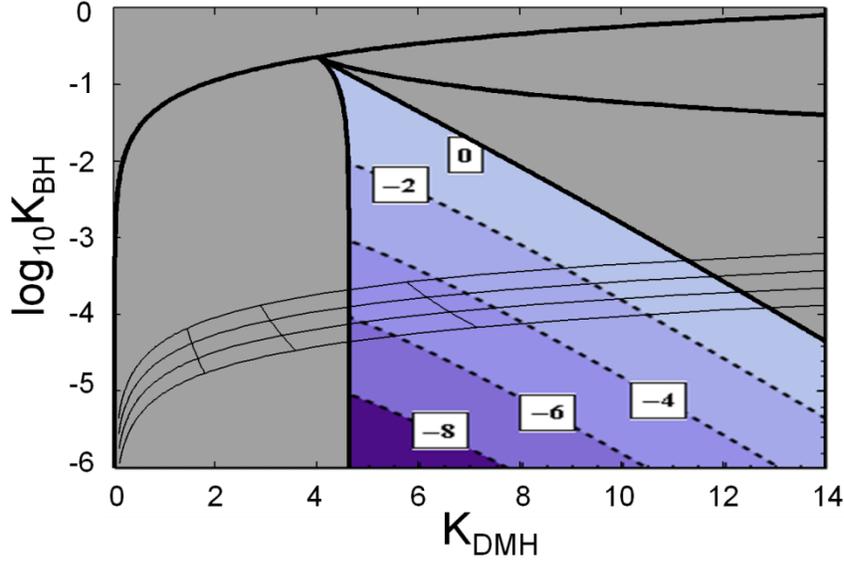

**FIGURE 2.** The contours represents the ratio of mass fluxes $\log_{10}\xi$ (defined in Equation (10)). The vertical axis is $\log_{10}K_{BH}$ and the horizontal axis is $K_{DMH}$.

We estimate the amount of the metals ejected by these outflows as follows,

$$\begin{aligned}\dot{Z}_{X_{in}} &= y_{X_{in}}\dot{M}_{X_{in}} \\ \dot{Z}_{X_{out}} &= y_{X_{out}}\dot{M}_{X_{out}}\end{aligned} \Rightarrow \frac{\dot{Z}_{X_{in}}}{\dot{Z}_{X_{out}}} = \frac{y_{X_{in}}}{y_{X_{out}}}\frac{\dot{M}_{X_{in}}}{\dot{M}_{X_{out}}} = \frac{y_{X_{in}}}{y_{X_{out}}}\xi \quad \ldots (11)$$

where $y$ is yield of metals and $\dot{Z}$ is metal mass ejected by outflows. The yield $y$ depends on the locus of the starting point of outflow. Assuming the metallicity gradient in elliptical galaxies[9)]

$$\frac{d[Fe/H]}{d\ln r} = -0.2, \quad \ldots (12)$$

we obtain $y_{X_{in}}/y_{X_{out}} \approx 2-3$.

In the Sombrero galaxy, temperature $T = 0.6 keV$, the mass of DMH $M_{DMH} = 10^{13} M_\odot$ and the SMBH mass $M_{BH} = 10^9 M_\odot$ [10,11,12]. Using these parameters, $\dot{M}_{X_{in}} = 0.006 M_\odot yr^{-1}$ and $\dot{M}_{X_{out}} = 9.97 M_\odot yr^{-1}$. The metallicity $[Fe/H]$ reaches a peak at -1.4 (<25kpc)[13]. For $X_{out}$, the total ejected metal mass is estimated $7.9 \times 10^7 M_\odot$ for 10Gyr. This outflow could pollute the intergalactic space to a mean metallicity $[Z] = -3$ (comparable to the levels observed in the Lyα forest at z=3)[14]. This result indicates the possibility of strong influence of slowly accelerated outflows ($X_{out}$) on the metal enrichment of the intergalactic space.

## SUMMARY

We have categorized possible transonic solutions of galactic outflows in the gravitational potential of DMH and SMBH using the isothermal, spherically symmetric and steady model. We conclude that the gravitational potential of SMBH generates a new transonic branch while Tsuchiya et al. (2013) concluded that the gravitational potential of DMH forms one transonic solution. Because these two transonic solutions have different mass fluxes and starting points, these solutions will make different influences to the star formation rate, the evolution of galaxies, and the chemical evolution of the intergalactic medium. Therefore, we conclude that the influence of galactic outflows to the intergalactic medium depends not only on the mass distribution but also on the selected transonic solution. In addition, we have estimated the range of parameters ($K_{DMH}$; $K_{BH}$) for actual galaxies.

Moreover, it may be possible to estimate the galactic mass distributions of DMH and SMBH applying the model to the observed profile of the outflow velocity. Although it is difficult to determine the velocity of hot gas in the galactic halos from the current X-ray observations, but the next-generation X-ray observatory will be able to detect the detailed profiles of outflow velocities.